\documentclass[aps,superscriptaddress,twocolumn]{revtex4}

\usepackage{graphicx}
\usepackage{dcolumn}
\usepackage{bm}
\usepackage{amssymb,amsthm}
\usepackage{lineno}
\usepackage{subfigure}
\usepackage{textcomp}
\usepackage{booktabs}
\usepackage{xcolor}

\begin{document}
\title{Assessment on Thermal Transport Properties of Group-III Nitrides: A Classical Molecular Dynamics Study with Transferable Tersoff Type Inter-atomic Potentials}

\author{Yenal Karaaslan}
\email{yenalkaraaslan@gmail.com}
\affiliation{Department of Mechanical Engineering, Eski\c{s}ehir Technical University, 26555, Eski\c{s}ehir, Turkey}
\author{Haluk Yapicioglu}
\email{hyapicio@eskisehir.edu.tr}
\affiliation{Department of Industrial Engineering, Eski\c{s}ehir Technical University, 26555, Eski\c{s}ehir, Turkey}
\author{Cem Sevik}
\email{csevik@eskisehir.edu.tr}
\affiliation{Department of Mechanical Engineering, Eski\c{s}ehir Technical University, 26555, Eski\c{s}ehir, Turkey}
\date{\today}

\begin{abstract}
In this study, by means of classical molecular dynamics simulations, we investigated the thermal transport properties of hexagonal single-layer, zinc-blend and wurtzite phases of BN, AlN, and GaN crystals, which are very promising for the application and design of high-quality electronic devices. With this in mind, we generated fully transferable Tersoff-type empirical inter-atomic potential parameter sets by utilizing an optimization procedure based on particle swarm optimization. The predicted thermal properties as well as the structural, mechanical and vibrational properties of all materials are in very good agreement with existing experimental and first-principles data. The impact of isotopes on thermal transport is also investigated and between $\sim$10 and 50\% reduction in phonon thermal transport with random isotope distribution is observed in BN and GaN crystals. Our investigation distinctly shows that the generated parameter sets are fully transferable and very useful in exploring the thermal properties of systems containing these nitrides.
\end{abstract}

\maketitle

\section{\label{sec:level1}Introduction}
Over the past three decades, the binary group III-nitride semiconductors, in particular GaN, AlN, and their alloys have attracted extraordinary amount of interest due to their usability in electronic and optoelectronic device applications, requiring high-efficiency at short wavelengths, high operation temperatures, high powers, and high frequencies~\cite{Strite,Ambacher,Vurgaftman,Vurgaftman1,Nakamura,Ferreyra}.
Therefore, these materials have been studied extensively and their superior physical properties such as wide band-gap, high thermal conductivity, short bond length, high dielectric constant, and low compressibility has been emphasized as the prominent characteristics of these crystals~\cite{Miwa,Gorczyca,Christensen,Kwiseon,Shul,Shimada,Przhevalskii,Junqiao}.
These peculiar properties have provided this material family with various application domains such as high electron mobility transistors, laser diodes, light-emitting diodes, photo-detectors, solar cells, electro-optic modulators, and biosensors~\cite{Nakamura1,Taniyasu,Mokkapati,Lu,Jos,Li}.
Typically, these extraordinary materials grow in the $wurtzite$ crystal structure where atoms are fourfold coordinated adopting $sp^{3}$-hybridization. However, with the recent advances in fabrication technologies, the low-dimensional graphene-like (threefold coordinated adopting $sp^{2}$-hybridization) structures of  BN~\cite{Tsai,Song,Nakhaie,Tonkikh,Vuong,Cheng}, AlN~\cite{Tsipas,Mansurov}, and GaN~\cite{Seo,Balushi} have been successfully fabricated with high dimensional accuracy.
And research studies on the investigation of the capability of these materials with regard to future device applications, particularly to address desired factors of high computing performance, low power consumption, cool operation, and lightweight have begun to appear in the literature~\cite{Sahin,Guo,Huang,Kecik}. 

Similar to their bulk counterparts, adopting this material family for future device technologies is quite probable.
Therefore, the accurate characterization of particularly device-related physical properties of these crystals is of utmost importance. For instance, thermal transport properties that provide a basis for thermal energy control and thermal management of electronic and optoelectronic devices needs to be systematically investigated.
In fact, efficient thermal management is required to cope with the problem of excess heat that occurs in conjunction with the efforts to miniaturize devices and improve their performance parameters. Indeed, this is directly related to the understanding of thermal transport properties of materials used in device fabrication. 

One of the best approaches to investigate the thermal transport properties of materials regarding device applications is classical molecular dynamics (CMD) simulations due to the fact that it is suitable to investigate large-scale systems in the order of millions of atoms. 
CMD simulations have proven to generate highly accurate results for thermal transport properties of bulk and nano structures, including disorders such as grain boundary, vacancy, and isotope defects~\cite{Fennimore,Broido,Lindsay,Thomas,Tabarraei}.
However, CMD simulations require an accurate empirical inter-atomic potential (EIAP) generated specifically aiming at the desired physical properties. In the case of thermal transport properties, accurate description of atomic vibrations is essential.
Also, transferability of the generated EIAP is highly critical to investigate  material considered in its different crystal structures. An accurate transferable EIAP is crucial not only in the development of heat control mechanisms for electronic device applications such as information,  communication, and energy storage technologies~\cite{Yuanpeng2017,Feng2018,Song2018}, but also in nanostructure device application simulations, such as drug delivery, superlubricity, and thermal rectification\cite{PhysRevLett.122.076102,C5RA22945F,Chen_2018}. 

\begin{figure}[!ht]
\includegraphics[width=\linewidth]{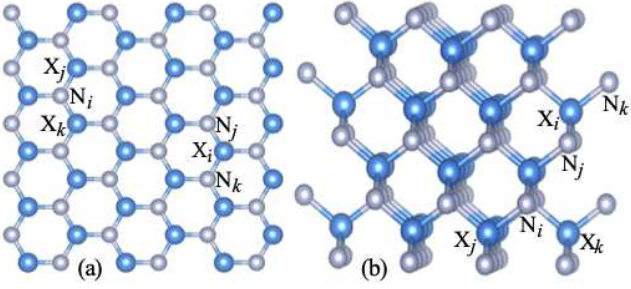}
\caption{(color online) Schematic representation of the group-III nitride crystal structures (X=boron, aluminum, gallium; N=nitrogen; $i$ is the central atom, $j$ and $k$ are two neighbor atoms bonded to the central atom).
(a) The top view for the hexagonal single-layer structure, and (b) the cross view for the zinc-blend bulk structure.}
\label{fig1}
\end{figure}

In this study, we first optimized a Tersoff type EIAP parameter set for binary group-III nitride compounds, BN, AlN, and GaN and we systematically investigated lattice thermal transport properties of these materials.
The generated parameter sets for each material were proven to accurately describe structural, mechanical, dynamical as well as thermal transport properties. In addition, considering different three-body parameter sets, i.e. B-N-N, and N-B-B, vacancy defect energies were generated in high accuracy. 
The transferability of the generated parameters was demonstrated by systematically testing on hexagonal single-layer, zinc-blend and wurtzite structures.
Our results indicated that isotope disorder has a strong influence on thermal transport properties of BN and GaN crystals.

\section{\label{sec:level2} Computational methods and details}

The form of considered three-body Tersoff inter-atomic potential~\cite{Tersoff} for the energy, $E$ of a system of atoms  is expressed as follows:
\begin{eqnarray}\label{Tersoff}
E &=& \frac{1}{2}\sum_{i}\sum_{j \neq i}V_{ij}
\\\nonumber
V_{ij} &=& f^C(r_{ij}) \left[a_{ij} f^R(r_{ij}) -  b_{ij} f^A(r_{ij}) \right]\;,
\\\nonumber
    f^C(r) &=& \left\{\begin{array}{lr}
        1\;, &  r < R-D\\
        \frac{1}{2} - \frac{1}{2} \sin\left[\frac{\pi}{2} \frac{(r-R)}{D} \right], &  R-D < r < R+D\\
        0\;, &  r < R+D\;
        \end{array}\right.
        \\\nonumber
f^R(r) &=& A \exp(-\lambda_1 r),
\\\nonumber
f^A(r) &=& B \exp(-\lambda_2 r),
\\\nonumber
a_{ij} &=& \left(1 + \alpha^n \eta_{ij}^n \right)^{-1/2n}
\\\nonumber
b_{ij} &=& \left(1 + \beta^n \xi_{ij}^n \right)^{-1/2n}
\\
\nonumber
\xi_{ij} &=& \sum_{k\neq i,j} f^C(r_{ik}) g(\theta_{ijk}) \exp\left[\lambda_3^3 (r_{ij} - r_{ik})^3\right]\\
g(\theta_{ijk}) &=& 1 + \frac{c^2}{d^2} - \frac{c^2}{d^2 + (h-\cos\theta_{ijk})^2} \;.
\end{eqnarray}
The summations in the formula are over all neighbors $j$ and $k$ of atom $i$ within a cut-off distance, $R$ + $D$. Here, $r_{ij}$ is the distance between atoms $i$ and $j$, $f^R$ is the repulsive potential energy function, $f^A$ is the attractive potential energy function, and $f^C$ is a smooth cut-off function that limits the range of the potential over the nearest neighbor interactions. The $a_{ij}$ and $b_{ij}$ are many-body order parameters that determine the effect of atomic arrangements of neighboring atoms on the energy of the system. The $g(\theta_{ijk})$ is the bond angle between $i-j$ and $i-k$ pairs, as described in Fig.~\ref{fig1}. 

In this study, as recommended by Tersoff the term, $a_{ij}$ was set to 1 ($\alpha = 0$). Therefore, four two-body terms, $A$, $B$, $\lambda_1$, $\lambda_2$ and six three-body terms $\lambda_3$, $n$, $\beta$, $c$, $d$, $h$ were considered to be parameters values of which can be engineered to obtain desired physical characteristics. Also, in order to accurately describe the three-body interactions, particularly for the correct description of vacancy defect formation energies, X-N-N and N-X-X three-body interaction parameters were generated separately (first one is the center atom, second and third are the atoms bonded to the center atom).      

The particle swarm optimization (PSO) algorithm as explained in our previous study~\cite{Kandemir} was used to generate the described 16 EIAP parameters for each material. For this purpose the fitness function for the PSO was defined as:
\begin{widetext}
\begin{eqnarray}\label{ObjectiveFunction}
f (A, B, \lambda_1, \lambda_2, \lambda_3^X, n^X, \beta^X, c^X, d^X, h^X, \lambda_3^N, n^N, \beta^N, c^N, d^N, h^N) = \sum_{j=1}^{J} \left( \frac{|d_j - a_j|}{d_j} \right).
\end{eqnarray}
\end{widetext}
Where, $d_j$ denotes the desired value of the characteristic $j$ obtained by first-principles calculations, $a_j$ denotes the actual value of the characteristic $j$ obtained via empirical potential for a given set of parameters, and $J=58$ is the total number of the certain physical characteristics of two crystal phases (hexagonal monolayers, $h$-BN, $h$-AlN, and $h$-GaN and zinc-blend bulk structures, $zb$-BN, $zb$-AlN, and $zb$-GaN) to be optimized simultaneously. Here, the lattice constant ($a_0$), phonon frequencies ($\omega$) corresponding to the selected acoustic and optic vibrations with different wavelengths, equation of states (EOS) defined as the deviation from the equilibrium energy via isotropic tensile and compressive strain, the formation energy difference between the $h$ and $zb$ crystal phases, and the formation energies corresponding to five different vacancy defect structures shown in Supplementary Materials (SM) were considered. The values were determined by using the General Utility Lattice Program (GULP) code~\cite{Gulp} throughout the EIAP parameter set optimization process.

The desired values of the variables were calculated via first-principles pseudopotential plane-wave simulations based on the density functional theory (DFT) using the Vienna \textit{ab initio} simulation package (VASP)~\cite{Kresse,Blochl,Kresse2}. In order to minimize the periodic layer interactions of the hexagonal structures, a vacuum spacing of $15~$\r{A} along the $z$-direction was considered. A plane wave basis set with $600$ ($700$)~eV kinetic energy cut-off and the $\Gamma$ point centered $24\times24\times1$ ($12\times12\times12$) $k$-point mesh within the Monkhorst-Pack scheme for the Brillouin zone integration of the primitive cell were used for all the hexagonal monolayer (zinc-blend bulk) structures. In order to obtain the vibrational frequencies, the phonopy~\cite{Phonopy} code was employed by using the force constants computed from the density functional perturbation theory~\cite{Baroni} by means of VASP. For all the monolayer (bulk) structures, these calculations were carried out with $4\times4\times1$ ($3\times3\times3$) conventional supercell structure, considering the $\Gamma$ centered $6\times6\times1$ ($4\times4\times4$) $k$-point grids for the Brillouin zone sampling. The monovacancy, divacancy and Stone-Wales defect formation energies were calculated as:
\begin{eqnarray}\label{DFTDefect}
E_f^{DFT} = E_d - E_p + x E_X + n E_N\;,
\end{eqnarray}
where $E_p$, $E_d$, $E_X$ and $E_N$ are the total energy of the perfect crystal structure, the energy of structure with defect, the ground state energy of X elements, and the ground state energy of N, respectively. $x$ and $n$ stands for the number of missing X, and N atoms, respectively.

The transport coefficients were calculated through CMD by using the Green-Kubo relations derived from the fluctuation dissipation theorem\cite{McGaughey}. In Green-Kubo method, the thermal conductivity that relates to the ensemble average of the heat current auto-correlation function (HCACF) is given by
\begin{eqnarray}\label{GreenKubo}
\kappa_{\alpha\alpha} = \frac{1}{V k_B T^2} \int_0^\infty \langle J_{\alpha}(0) J_{\alpha}(t) \rangle dt
\end{eqnarray}
where $\alpha$ represent the three Cartesian coordinates ($x$-, $y$-, and $z$-directions), $V$ is the volume of the simulation cell, $k_B$ is the Boltzmann constant, $T$ is the temperature of the system,
and $J_{\alpha}(t)$ is the heat current calculated as follows~\cite{Mortazavi}: 
\begin{eqnarray}\label{HeatCurrent}
\textbf{J} = \sum_{i} \left( E_i \textbf{v}_i + \frac{1}{2} \sum_{i<j} \left(\textbf{F}_{ij} \cdot (\textbf{v}_i + \textbf{v}_j) \right) \textbf{r}_{ij} \right),
\end{eqnarray}
where $E_i$ is the total energy of the atom $i$, $\textbf{v}_i$ is the velocity of atom $i$, $\textbf{r}_{ij}$ denotes the inter-atomic distance between atoms $i$ and $j$, and $\textbf{F}_{ij}$ stands for the inter-atomic force.

The CMD simulations for thermal transport analysis were performed using the Large-scale Atomic/Molecular Massively Parallel Simulator (LAMMPS)~\cite{Plimpton,Lammps}, with $1\times10^6$ time steps ($\Delta t$ = 0.5~fs) in canonical ensemble (NVT) to reach the thermal equilibrium, and another $1\times10^7$ time steps in microcanonical ensemble (NVE) for heat current calculations. Then, the mean HCACF,  $(J_{\alpha}(0) J_{\alpha}(t))_k$ was obtained using the calculated heat current data by considering 500.000 time steps and 9.5$\times$10$^6$ initial point, $k$ with lags of increments of 10 time steps. In addition, the whole procedure was repeated for ten different sets of initial particle velocities randomly distributed by a Gaussian distribution as defined in LAMMPS. Next, averaging over 10 different trials provide us with the overall average of HCACF, and the resulting figures are provided in the SMs as examples. Finally, the data obtained as explained above were used to compute final lattice thermal conductivity with the following equations~\cite{McGaughey,Nichenko}:  
\begin{eqnarray}\label{RelaxTime}
& & \langle \textbf{J}(0) \cdot \textbf{J}(t) \rangle = A_1 e^{-t/\tau_1} + A_2 e^{-t/\tau_2}. \nonumber\\
& & \kappa = \frac{1}{V k_B T^2} \left(A_1 \tau_1 + A_2 \tau_2\right).\nonumber
\end{eqnarray}
Here, $\tau_{1}$ and $\tau_{2}$ represent time constants, $A_1$ and $A_2$ represent the strength of phonon modes. The basic logic in the definition of two variables is to distinguish the contribution of short-range optical and long-range acoustic modes, which can make significant differences in thermal conductivity for some materials.

The CMD simulations for single layer hexagonal structures were performed in quasi square simulation cells, width of which is $\sim$81, $\sim$102, and $\sim$106~nm for, BN, AlN, and GaN, respectively. And for zinc-blend and wurtzite structures simulations were performed in cubic cells with the edge length of $\sim$16, $\sim$19, and $\sim$20~nm for, BN, AlN, and GaN, respectively. In the volume calculation of the single layer structures, we assumed an effective layer thickness of $3.33~$\r{A} for monolayer $h$-BN, $3.4~$\r{A} for monolayer $h$-AlN, and $3.49~$\r{A} for monolayer $h$-GaN, in accordance with the results we obtained from the first-principle calculations of hexagonal bulk structures. In order to factor in the isotopic disorder nature of the crystals, the simulation cell structures contain randomly distributed $20\%$ $^{10}$B and $80\%$ $^{11}$B isotopes for BN, and  $60\%$ $^{69}$Ga and $40\%$ $^{71}$Ga isotopes for GaN.
\section{\label{sec:level3}Results}
\begin {table*}[!ht]
\caption{Optimized Tersoff-type empirical inter-atomic potential parameters obtained by using the PSO method for the BN, AlN, and GaN.}\label{tableTersoff} 
\begin{center}
\begin{tabular}{lllllll}
\toprule[1.5pt]
 & B-N & N-B & Al-N & N-Al & Ga-N & N-Ga \\\toprule[1.1pt]
$A$ ($eV$)& 1205.446293 & 1205.446293 & 1258.567263 & 1258.567263 & 2249.391746 & 2249.391746\\\midrule
$B$ ($eV$)& 436.750025 & 436.750025 & 453.228512 & 453.228512 & 764.751142 & 764.751142\\\midrule
$\lambda_1$ (\r{A}$^{-1}$) & 2.965635 & 2.965635 & 2.434869 & 2.434869 & 2.652624 & 2.652624\\\midrule
$\lambda_2$ (\r{A}$^{-1}$) & 2.060658 & 2.060658 & 1.717680 & 1.717680 & 1.963739 & 1.963739\\\midrule
$\lambda_3$ (\r{A}$^{-1}$) & 1.165721 & 1.108293 & 1.186759 & 1.100709 & 1.453060 & 1.166408\\\midrule
$n$ & 1.156834 & 0.939775 & 0.598233 & 1.220882 & 0.761872 & 1.070552\\\midrule
$\beta$ & 0.741928e-6 & 1.335243e-6 & 2.133047e-6 & 2.496023e-6 & 2.143554e-6 & 2.642801e-6\\\midrule
$c$ & 26320.215836 & 27483.938392 & 19110.741778 & 27568.039128 & 45996.528912 & 44736.208668\\\midrule
$d$ & 5.255691 & 7.783163 & 10.882090 & 8.292436 & 13.083985 & 12.497037\\\midrule
$h$ & -0.870210 & -0.749741 & -0.972662 & -0.816226 & -0.990019 & -0.693793\\\midrule
$R$ (\r{A})& 2.20 & 2.20 & 2.55 & 2.55 & 2.65 & 2.65\\\midrule
$D$ (\r{A})& 0.10 & 0.10 & 0.15 & 0.15 & 0.15 & 0.15\\
\bottomrule[1.5pt]
\end{tabular}
\end{center}
\end {table*}

As previously mentioned, lattice constants, equation of states, phonon frequencies, defect energies and the difference of formation energies per atom of single layer $h$ and bulk $zb$ crystal phases are included as physical characteristics in the potential fitting procedure. The Tersoff type EIAP sets obtained using PSO are presented in Table~\ref{tableTersoff} for BN, AlN and GaN. In SM file, the comparative results for all the physical quantities considered (first-principle results and the values obtained with the optimized potential parameters) are presented. The results show that the generated transferable EIAP set  clearly represent the structural, mechanical, and dynamical properties of both $h$ and $zb$ structures of considered materials within a reasonable margin of error. 

Due to many possible unanticipated effects that may occur in the experimental fabrication procedures in different growth methods, yielding a material free from defect formations is highly improbable~\cite{Egerton,Chuanhong}, in particular for monolayer structures~\cite{Chuanhong,Jannik,Nasim}. Therefore, in order to obtain a parameter set well describing the characteristic of considered materials in the presence of defects, the mono-vacancy, bi-vacancy and Stone-Wales defect~\cite{Dmitry,Shahriar} formation energies are factored in during the optimization procedure. The comparative results of the formation energies of five different vacancy defect structures created for the hexagonal single-layer phases of each material are presented in SMs. The listed energies are in very close agreement with first-principles calculations except for $V_X$ and $V_{X+3N}$ defects. Our definition of three-body parameter clearly worked well in distinguishing the $V_X$ and $V_N$ mono-vacancy formation energies. 

\begin{figure*}[!ht]
\includegraphics[width=\linewidth]{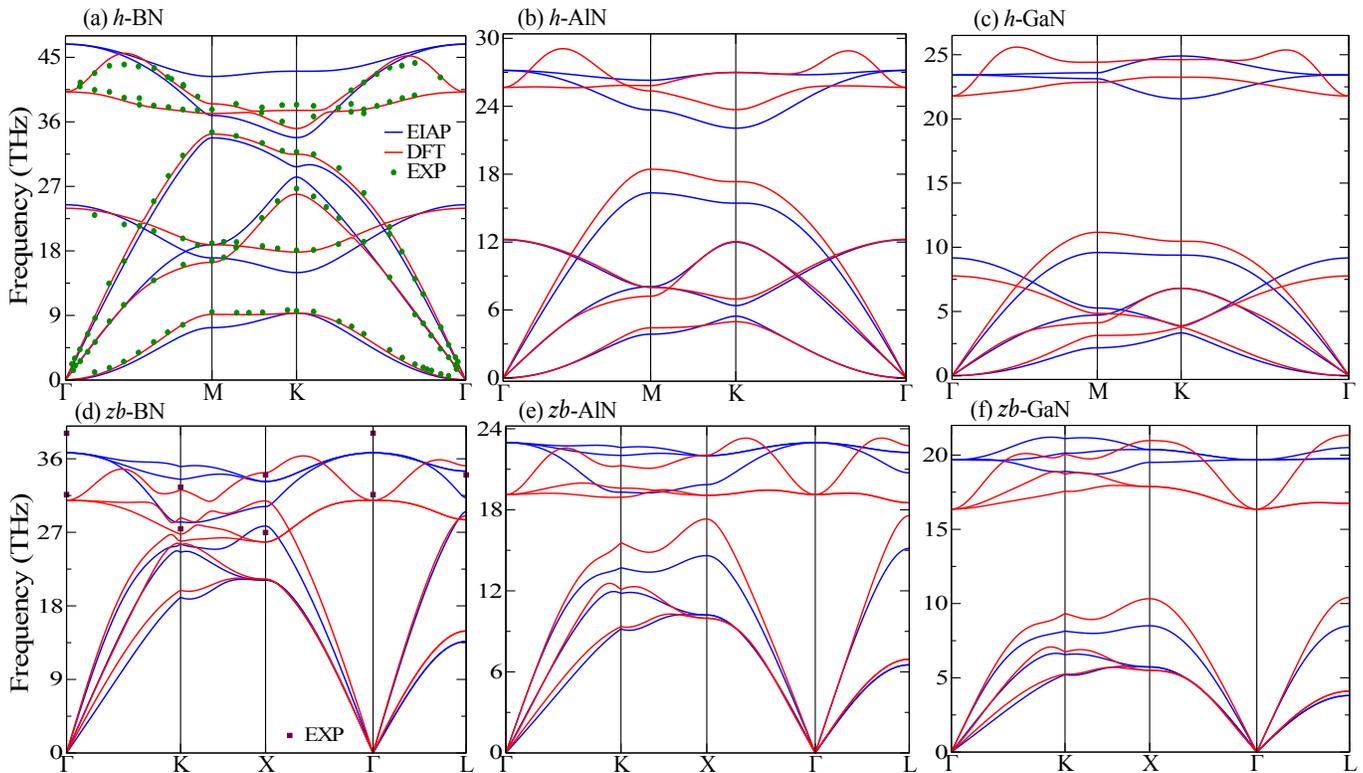}
\caption{(color online) Phonon dispersions of the hexagonal monolayer (a) BN,  (b) AlN, (c) GaN, and zinc-blende bulk (d) BN, (e) AlN, and (f) GaN along the high-symmetry reciprocal space points. The results of the first-principles (DFT, red line), force-field based calculations (EIAP, blue line) and experimental data (green circle from Ref.~\cite{PhysRevLett.98.095503}, maroon square from Ref.~\cite{PhysRevB.71.205201}).}
\label{figPhonon}
\end{figure*}

Phonons are the primary heat carriers in semiconductors~\cite{Xiaokun}. In order to fully grasp the thermal dependent properties of a material, an accurate characterization of the vibrational properties is essential. The phonon dispersion curves obtained with EIAP sets are substantially compatible with the first-principle results as shown in Fig.~\ref{figPhonon} (a)-(f). In addition, experimental data for $h$-BN~\cite{PhysRevLett.98.095503} and $zb$-BN~\cite{PhysRevB.71.205201} structures have been added to phonon dispersions, and it is observed that the results of force-field based calculations are in good agreement with measurement results. Notably, acoustic branches have a great coherence with the first-principle calculations for both monolayer and bulk phases of the BN, AlN and GaN. In the first-principles phonon calculations we do not include longitudinal and transverse optic modes splitting. Since we try to produce pair potential parameters, calculated errors in optic phonons is acceptable up to the deviation of the longitudinal-transverse optic splitting effect. The noticeable difference in high-lying optic modes can be considered as the weakness of generated parameter set, however, the effect of these modes on lattice thermal conductivity is relatively less than the acoustic modes due to their low phonon group velocities~\cite{PhysRevB.97.104308,MAHDIZADEH20161}. On the other hand, one should consider the in-direct effect of optical modes due to phonon-phonon scattering. The results clearly show that optical phonon representation of our EIAP parameter sets are good enough to get reasonable results for thermal properties. Consequently, one can clearly conclude taking into account the represented comparisons that the generated pairwise transferable inter-atomic parameter sets accurately describe the desired physical properties of group-III nitrides. 

\begin{figure}[!ht]
\includegraphics[width=\linewidth]{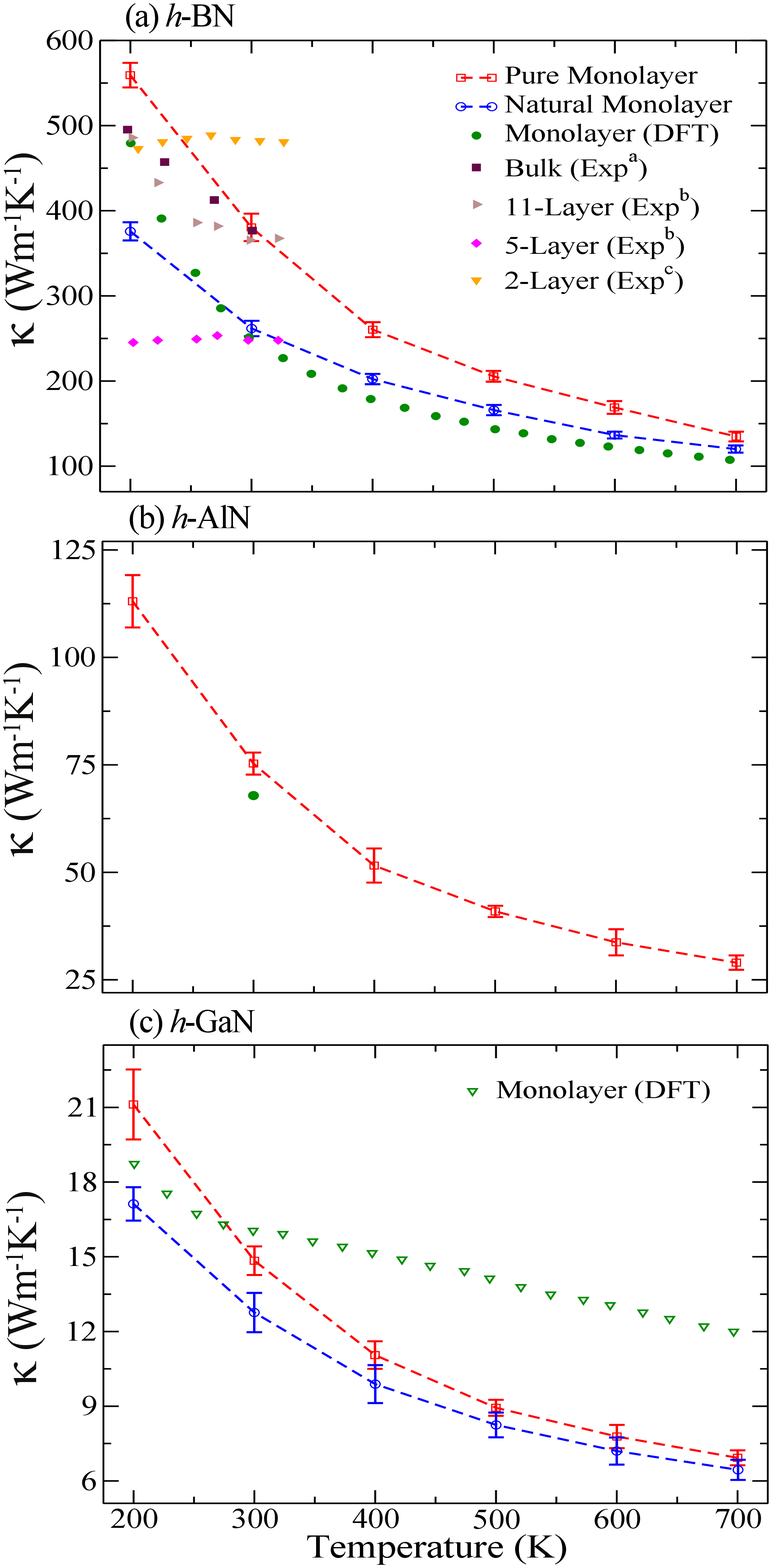}
\caption{(color online) The calculated lattice thermal conductivity, $\kappa$ for the isotopically pure (red dashed lines) and natural (blue dashed lines) hexagonal monolayer (a) BN,  (b) AlN, (c) GaN crystals as a function of temperature. Theoretical literature data: monolayer (DFT)  from Ref.~\cite{Guangzhao} for BN and AlN, from Ref.~\cite{PhysRevB.95.195416} for GaN (for this data, effective layer thicknesses are normalized according to our study). Experimental data: a, b, c from Ref.~\cite{Sichel,Insun,Chengru}, respectively.}
\label{figTC}
\end{figure}

Subsequent to the potential validation calculations, we predicted the lattice thermal conductivity, $\kappa=(\kappa_{xx}+\kappa_{yy})/2$ of the monolayer BN, AlN, and GaN structures in the $200-700$ ~K temperature range as shown in Fig.\ref{figTC}. The  $\kappa$ of BN is observed to decrease from $\sim$560 to $\sim$120~Wm$^{-1}$K$^{-1}$ between the 200 and 700~K, and  room temperature value is calculated as $380$~Wm$^{-1}$K$^{-1}$. These results are quite comparable with our previous prediction by non-transferable Tersoff potential~\cite{PhysRevB.84.085409}. We also investigated the effect of isotope disorder on $\kappa$ and found out that the room temperature $\kappa$ of natural (with $20\%$ $^{10}$B and $80\%$ $^{11}$B) BN is obtained as 260~Wm$^{-1}$K$^{-1}$. The $\sim$30\% decrease on $\kappa$ shows the strong influence of isotope disorder on thermal transport properties of the material as previously predicted~\cite{Guangzhao,PhysRevB.86.075403}. The room temperature thermal conductivity for bulk $h$-BN was reported around 390~Wm$^{-1}$K$^{-1}$ by Sichel \textit{et al.}~\cite{Sichel}. Also, the room temperature thermal conductivity values were calculated around 360~Wm$^{-1}$K$^{-1}$ for 11-layers $h$-BN~\cite{Insun}, 227-280~Wm$^{-1}$K$^{-1}$ for 9-layers $h$-BN~\cite{Haiqing}, 250~Wm$^{-1}$K$^{-1}$ for 5-layers $h$-BN~\cite{Insun}, and 484~Wm$^{-1}$K$^{-1}$ for 2-layers $h$-BN~\cite{Chengru}. Moreover, Cai \textit{et al.} recorded the thermal conductivity of 1, 2, and 3-layers $h$-BN as 751, 646 and 602~Wm$^{-1}$K$^{-1}$, respectively, at close-to room temperature using optothermal Raman measurements~\cite{Cai}. Despite the notable deviation among the reported results in the literature, our results are in reasonable agreement with both experimental and theoretical calculations. Indeed, the underestimation of CMD simulation on $\kappa$ is a well known fact due to the collective excitation of phonon modes even at low temperatures. 

For $h$-AlN, $\kappa$ decreases from $\sim$115 to $\sim$28~Wm$^{-1}$K$^{-1}$ within the same temperature range, while it is around 75~Wm$^{-1}$K$^{-1}$ at room temperature as seen in Fig.\ref{figTC} (b). Qin \textit{et al.}~\cite{Guangzhao} reported the room temperature value as 74.43~Wm$^{-1}$K$^{-1}$  by means of first-principles based solution of phonon Boltzmann Transport Equation (PBTE). Using the same thickness, we obtained $\kappa$ as 82~Wm$^{-1}$K$^{-1}$ which is in quite good agreement. 

The thermal conductivity of the $h$-GaN is observed to decrease from $\sim$21 to $\sim$6~Wm$^{-1}$K$^{-1}$ in the 200-700~K temperature range as shown in Fig.\ref{figTC} (c). The calculated room temperature value is predicted as 15~Wm$^{-1}$K$^{-1}$. This result is in conjunction with the first-principle based PBTE solution reported in the literature~\cite{Guangzhao,Zhenzhen,PhysRevB.95.195416,Jiang}, when the effective thickness values are selected in accordance with these studies. The percent abundance of Ga isotopes is as follows: $60\%$ $^{69}$Ga and $40\%$ $^{71}$Ga. Therefore, we investigated the effect of isotope disorder for also $\kappa$ of GaN. Our results clearly depicted that the effect of isotope disorder is around 10\%, mainly due to the fractional difference between the two isotope masses when compared with the $h$-BN. 

\begin{figure}[!ht]
\begin{center}
\includegraphics[width=\linewidth]{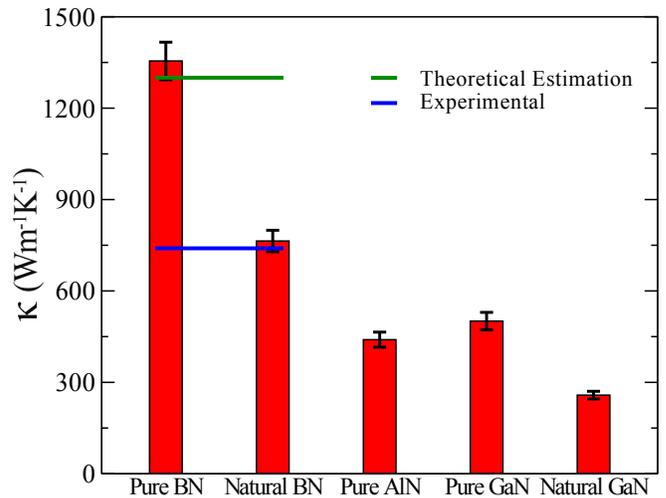}
\caption{(color online) The calculated room temperature lattice thermal conductivity, $\kappa$ values for the isotopically pure and natural zinc-blend bulk BN, AlN, and GaN crystals. Here, the experimental data is from Ref.~\cite{Michael}, the theoretical estimate is from Ref.~\cite{Slack}.}
\label{figTCBulk}
\end{center}
\end{figure}

The calculated room temperature lattice thermal conductivity ($\kappa=(\kappa_x+\kappa_y+\kappa_z)/3$) values for the $zb$ bulk phases are summarized in Fig.~\ref{figTCBulk}, as $\sim$1350, $\sim$440, and $\sim$501~Wm$^{-1}$K$^{-1}$ for $zb$-BN, $zb$-AlN, and $zb$-GaN, respectively. To the best of our knowledge, there is no experimental study on the thermal conductivity of the $zb$-AlN and $zb$-GaN in the literature. However, the measured room temperature $\kappa$ for $zb$-BN is 740~Wm$^{-1}$K$^{-1}$~\cite{Michael} which is almost half of the value calculated in this study, 1355~Wm$^{-1}$K$^{-1}$ and the reported theoretical estimation, $\sim$1300~Wm$^{-1}$K$^{-1}$~\cite{Slack,Michael}. But, when we factor in the isotope effect we practically get the same result as 764~Wm$^{-1}$K$^{-1}$. The reduction on the thermal conductivity with the isotope disorder is around 44\% for the $zb$-BN, and surprisingly 48\% for the $zb$-GaN.

\begin{figure*}[!ht]
\includegraphics[width=\linewidth]{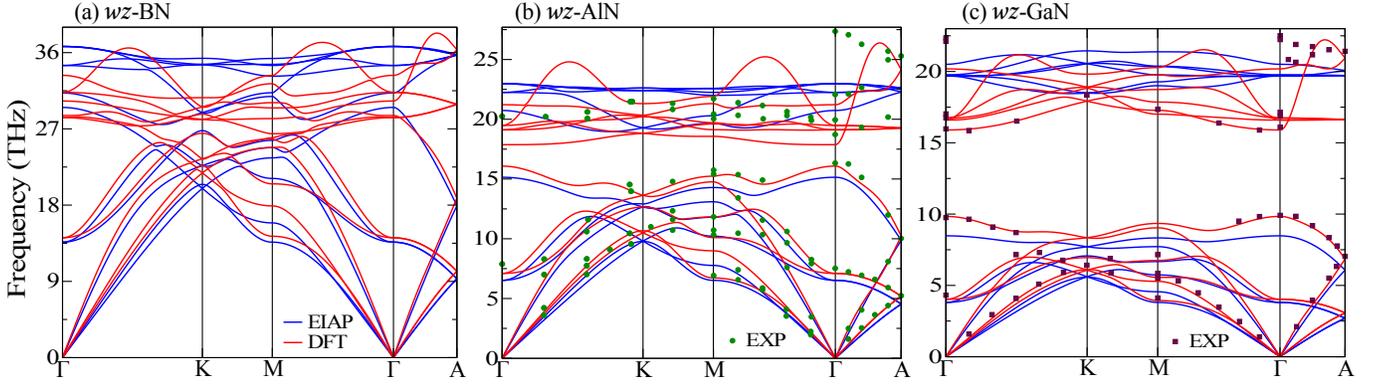}
\caption{(color online) Phonon frequencies for the $wz$ bulk (a) BN, (b) AlN, and (c) GaN along the high-symmetry directions of the Brillouin zone. The results of the first-principles (DFT, red line), force-field based calculations (EIAP, blue line) and experimental data (green circle from Ref.~\cite{Schwoerer}, maroon square from Ref.~\cite{PhysRevLett.86.906}).}
\label{figPhononWurtzite}
\end{figure*}

In addition to the crystal structures considered in the optimization process, we also tested the generated EIAP parameter sets on wurtzite ($wz$) crystal phases of the BN, AlN and GaN. The accurate description of the generated potential for the desired physical properties can be clearly verified from the values presented in the SM in comparison with first-principles calculations. For instance, the calculated lattice constants are in agreement with first-principles data within a 10\% margin of error. Also, the change in total energy via isotropic tensile and compression strain very well match with the first-principle calculation results throughout the entire workspace. Fig.~\ref{figPhononWurtzite} (a)-(c) presents the comparative phonon frequencies calculated with the DFT and EIAP parameter sets of the $wz$-BN, -AlN, and, -GaN structures along high-symmetry directions of the Brillouin zone. Also the measurement results for $wz$-AlN~\cite{Schwoerer} and -GaN~\cite{PhysRevLett.86.906} phases are presented in Fig.~\ref{figPhononWurtzite} (b)-(c). The results are quite consistent in particular for acoustic modes. Therefore, we can clearly claim that the transferable Tersoff potential parameters for the $wz$-BN, -AlN, and, -GaN structures are highly compatible with the results of the first-principle calculations, which is an important proof of the transferability of the EIAP parameters generated for the materials considered in this study.

\begin{figure}[!ht]
\begin{center}
\includegraphics[width=\linewidth]{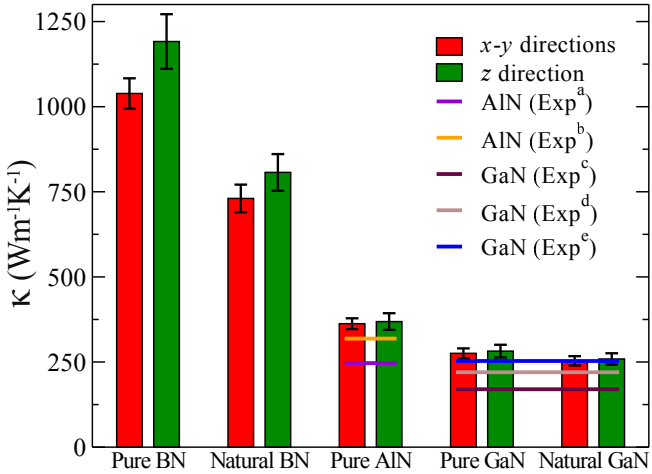}
\caption{(color online) The calculated room temperature lattice thermal conductivity values for the isotopically pure and natural wurtzite bulk BN, AlN, and GaN crystals. Here, experimental data a, b, c, d, e are from Ref.~\cite{doi:10.1063/1.5097172,Slack2,Pankove,Danilchenko,Shibata}, respectively.}
\label{figTCBulkWurtzite}
\end{center}
\end{figure}

The calculated lattice thermal conductivity for in plane, $\kappa_{ip} = (\kappa_x+\kappa_y)/2$ and out of plane, $\kappa_{op} = \kappa_z$ directions of $wz$ crystals are presented in Fig.~\ref{figTCBulkWurtzite}.
The room temperature values of $\kappa$ are about 1040 (1190), 360 (370), and 275 (280)~Wm$^{-1}$K$^{-1}$ for in plane (out of plane) for pure $wz$-BN, -AlN, and -GaN, respectively.
For $wz$-BN the results of the first-principles calculations recently reported by Chakraborty \textit{et al.}~\cite{Chakraborty}, $\kappa_{ip}$ ($\kappa_{op}$ )= 1344 (1155)~Wm$^{-1}$K$^{-1}$ are in parallel with our calculations. The measured $\kappa$ of $wz$-AlN at room temperature was experimentally~\cite{Slack,Slack2,Michael} estimated as 320~Wm$^{-1}$K$^{-1}$ (regardless of direction) and theoretically~\cite{Michael,Wu,Xuemei} reported as 285-400~Wm$^{-1}$K$^{-1}$ which are also consistent with our results. Recently, Xu \textit{et al.} have experimentally measured the thermal conductivity of $wz$-AlN by the $3w$ method, and have obtained $\kappa=237$ and $\kappa=247$~Wm$^{-1}$K$^{-1}$ at room temperature for two samples~\cite{doi:10.1063/1.5097172}. In addition, the resulting $\kappa_{ip}/\kappa_{op}$ ratio for AlN in our calculations, is 0.97 quite close to the value obtained by Li \textit{et al.}, as 0.95 (the average value is around 300~Wm$^{-1}$K$^{-1}$ ). Several different experimental~\cite{Slack,Pankove,Slack2,Florescu,Michael,Danilchenko,Shibata}  and theoretical~\cite{Lindsay,Zhenzhen}  studies report the average $\kappa$ of $wz$-GaN between 170 and 260~Wm$^{-1}$K$^{-1}$, and 260 and 410~Wm$^{-1}$K$^{-1}$, respectively. The calculated $\kappa$ value for GaN in this study is consistent with the reported results, however, the obtained $\kappa_{ip}/\kappa_{op}$ = 0.98 ratio is higher than those reported in several different theoretical methods by Qin \textit{et al.}~\cite{Zhenzhen}, in which $\kappa_{ip}/\kappa_{op}$ = 0.8-0.9. Another point to emphasize here is that the isotope disorder has a strong influence on thermal conductivity.
The effect of the isotope dispersion on thermal conductivity is about 30\% for $wz$-BN (730 and 807~Wm$^{-1}$K$^{-1}$ for in plane and out of plane directions, respectively) and 8\% for $wz$-GaN (253 and 259~Wm$^{-1}$K$^{-1}$ for in-plane and out of plane directions, respectively).

\section{\label{sec:level4}Conclusion}
In conclusion, we generated Tersoff-type transferable EIAP parameter sets for the BN, AlN and GaN crystals using a stochastic optimization algorithm, particle swarm optimization. The results clearly show that the generated parameters represent the structural, mechanical and dynamical nature of all the tested crystal phases of nitride compounds considered with an acceptable level of accuracy, even for the $wz$ crystals which are not included explicitly in the optimization procedure. Therefore, the generated fully transferable EIAP parameter sets can be adopted to investigate thermal properties of heterostructure and pristine bulk and nanosystems, even in the presence of isotope, vacancy and grain boundary type of defects. 

As we mentioned before, controlling and understanding energy dissipation and transport properties in nanostructure devices continue to be a rapid development and discovery area for more powerful, faster and smaller device applications. In this context, we test the generated EIAPs via a systematic investigation of thermal transport properties of well-known crystal structures of these compounds. Our results clearly demonstrate that our potential parameters reproduce the lattice thermal transport properties of these systems with a high level of agreement with both theoretical and experimental studies reported in the literature. In addition, we predicted a strong influence of isotope disorder on lattice thermal transport properties of BN and GaN crystals which clearly shows that disorder effects have to be taken into account in order to obtain more realistic results for materials and device systems.

\begin{acknowledgments}
The numerical calculations reported in this paper were partially performed at TUBITAK ULAKBIM, High Performance and Grid Computing Center (TRUBA resources). This work was supported by the Scientific and Technological Research Council of Turkey (TUBITAK), Grant No: MFAG-116F445. 
\end{acknowledgments}
\providecommand{\noopsort}[1]{}\providecommand{\singleletter}[1]{#1}%

\end{document}